\documentclass{aa}

\usepackage {graphicx}
\usepackage {longtable}
\def\etal{\emph{et al. }}
\begin{document}

\thesaurus{11     
            (   02.01.1 
                02.18.5 
                11.01.2 
                11.10.1 
                11.17.4 3C\,273)}
\title{The Jet of 3C\,273 observed with ROSAT HRI\thanks{Based on
observations with the ROSAT X-ray satellite and also on data collected with the
VLA (The National Radio Astronomy Observatory is a facility of the National
Science Foundation operated under cooperative agreement by Associated
Universities, Inc.) and at the European Southern Observatory, Chile (ESO
N$^\circ$ 51-2-021).}}
\titlerunning{3C\,273 with ROSAT HRI}
\author{Hermann-Josef R\"{o}ser\inst{1}, Klaus Meisenheimer\inst{1}, Martin Neumann\inst{1}
         \and R.G. Conway\inst{2} \and R.A. Perley\inst{3}}
\authorrunning{R\"{o}ser \emph{et al.}}
\offprints{H.-J. R\"{o}ser}
    \institute{Max-Planck-Institut f\"{u}r Astronomie, 69117 Heidelberg, Germany
    \and University of Manchester, Nuffield Radio Astronomy Laboratories, Jodrell Bank,
    Macclesfield, Cheshire SK11 9DL, U.K.
    \and National Rario Astronomy Observatory, P.O. Box 0, Soccoro, NM 87801, USA}
\date{Received March 31, 2000; accepted May 26, 2000}

\maketitle

\begin{abstract}
ROSAT HRI observations of 3C\,273 reveal X-ray emission all along the optically
visible jet with the peak of emission at the inner end. Whereas the X-ray
emission from the innermost knot\,A is consistent with a continuation of the
radio-to-optical synchrotron continuum, a second population of particles with
higher maximum energy has to be invoked to explain the X-ray emission from
knots B, C and D in terms of synchrotron radiation. Inverse Compton emission
could account for the X-ray flux from the hot spot. We detect a faint X-ray
halo with a characteristic scale of 29\,kpc and particle density on the order
of $6\times10^{4}$\,m$^{-3}$, higher than previously thought.

\keywords{quasars -- jets -- synchrotron radiation --- particle acceleration}
\end{abstract}

\section{Introduction}
Jets in extragalactic radio sources play a central r\^{o}le in our understanding
of the nature of these enigmatic sources (Begelman \etal\cite{BBR84}, R\"{o}ser
\etal\cite{RingbergII}) as they mark the channels through which mass, energy
and momentum are transported out from the nucleus into the extended radio
lobes. The detailed physical conditions in the jets are still unknown since
their synchrotron emission at radio frequencies provides little information
about the emitting plasma. However, of the more than 100 extragalactic radio
jets known (Bridle \& Perley \cite{BP84}) only three are readily detectable at
frequencies higher than the radio band. The two objects with the brightest
optical jet emission are M\,87, a radio galaxy, and 3C\,273, a quasar. Due to
this wide wavelength coverage we can expect that basic information about the
physical conditions giving rise to the synchrotron emission can be derived,
most importantly the maximum energy of the radiating particles.

We have therefore embarked on a detailed study of the jet of 3C\,273 at the
best angular resolutions currently available across the electromagnetic
spectrum using the VLA and MERLIN at radio wavelengths, HST at optical and
near-infrared and ROSAT at X-ray wavelengths (for a preliminary presentation of
some of these data see R\"{o}ser \etal\ \cite{HJRMNCDP97p}). Whereas the
synchrotron origin of the radio, infrared and optical emission is now firmly
established (Conway \& R\"{o}ser \cite{CHJR93}, R\"{o}ser \& Meisenheimer
\cite{HJRM91}) it is the origin of the X-rays that is least understood. The
most detailed X-ray study of this jet is due to Harris \& Stern
(\cite{HS87x}), who carefully analysed an EINSTEIN HRI observation with
95\,ksec integration time. Although they marginally detected a signal from the
jet, none of their attempts to interpret the X-ray emission proved
satisfactory. Our ROSAT observations were primarily aimed at verification and
interpretation of the jet's X-ray emission.

\section{The jet of 3C\,273 at radio to optical wavelengths}

\begin{figure*}[htbp]
\centering
\includegraphics[width=18cm,clip=true]{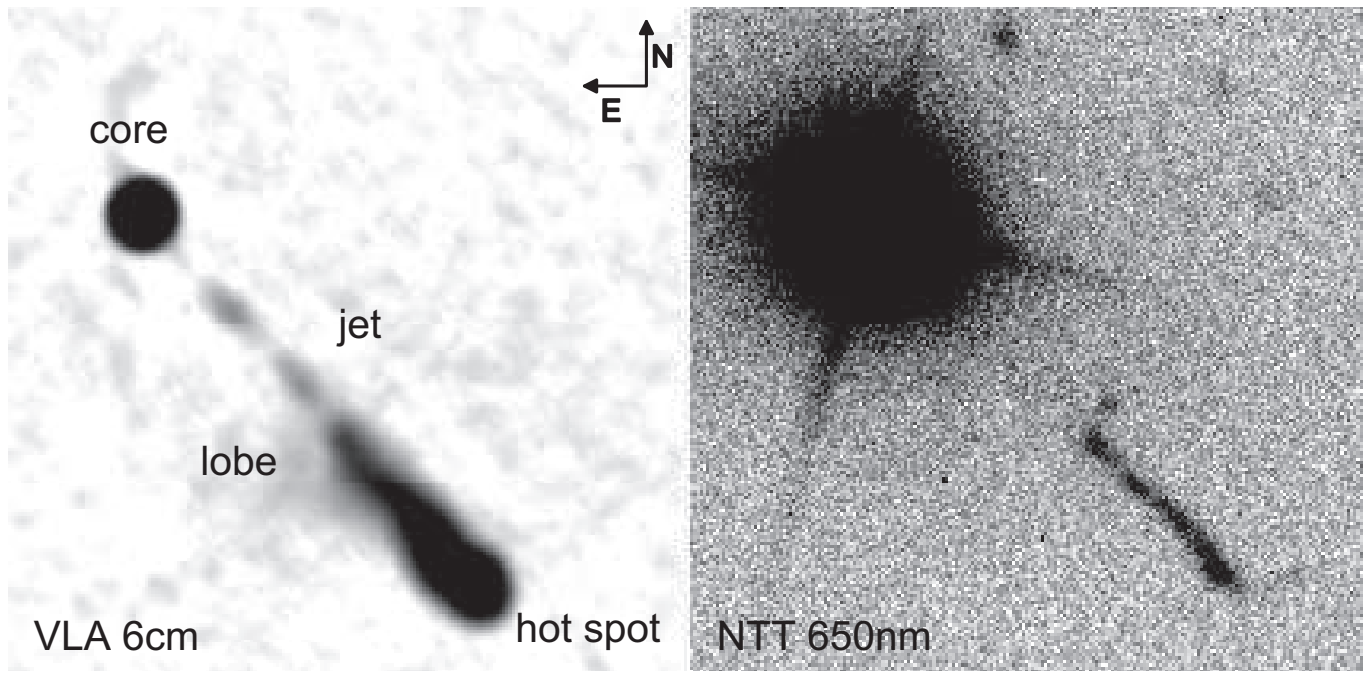}
\caption[]{\label{Fig1}The quasar 3C\,273 at radio (left) and optical (right)
wavelengths. The main morphological features are indicated. The radio jet
terminates in the bright radio hot spot at a distance of 21\farcs3 from the
core. The faint extensions visible to the north of both ends of the optical
jet are not detected at radio wavelengths.}
\end{figure*}

As indicated in Figure\,\ref{Fig1}, the radio jet is detected all the way from
the core out to the hot spot. The faint optical structure, however, although
coinciding in position angle with the line joining radio components A (hot
spot) and B (core), seems to be detectable only outwards of $ \approx
11$\arcsec. It also terminates about 1\arcsec\ before the peak in the radio hot
spot is reached. At the quasar's redshift of 0.158 the projected length is $
\approx $60\,kpc\footnote{We assume \emph{q}$_{0}$ = 0.1 and \emph{H}$_{0}$ =
65\,km/sec/Mpc, so 1\arcsec\ corresponds to 2.67\,kpc}. Greenstein \& Schmidt
(\cite{GS64}) in their detailed study of 3C\,273 and 3C\,48 briefly discuss
also this jet. Their spectrum of its outer end exhibited a featureless blue
continuum and they assumed the optical radiation is of the same origin as the
radio emission, \emph{i.e.} synchrotron radiation from relativistic particles.
This was proven by R\"{o}ser \& Meisenheimer (\cite{HJRM86}, \cite{HJRM91})
using optical polarimetry. Further hints about the synchrotron emission are
provided by studies of the spatially resolved continua of individual knots in
the jet. Meisenheimer \& Heavens (\cite{MH86}) present a simple model
describing the global shape of the continuum including an exponential cut-off
observed in the hot spot at the jet's outer end reflecting the maximum energy
gained by the relativistic particles. Applying this model to the other knots
in the jet indicates that we also see distributions of relativistic particles
truncated at some maximum energy, except for the innermost visible knot
(Meisenheimer \etal\ \cite{MNHJR96p}), which essentially has no cut-off at
all. In view of these results the marginal detection of X-ray emission by
Harris \& Stern (\cite{HS87x}) would naturally be associated with this
innermost knot, although they place the centroid of the X-ray emission further
out.

We have therefore used the ROSAT HRI with its better spatial resolution and
sensitivity to verify the X-ray emission from this jet and to gain further
insight into the emission mechanism.

\begin{figure}[htbp]
\centering
\includegraphics[width=8.8cm, clip=true]{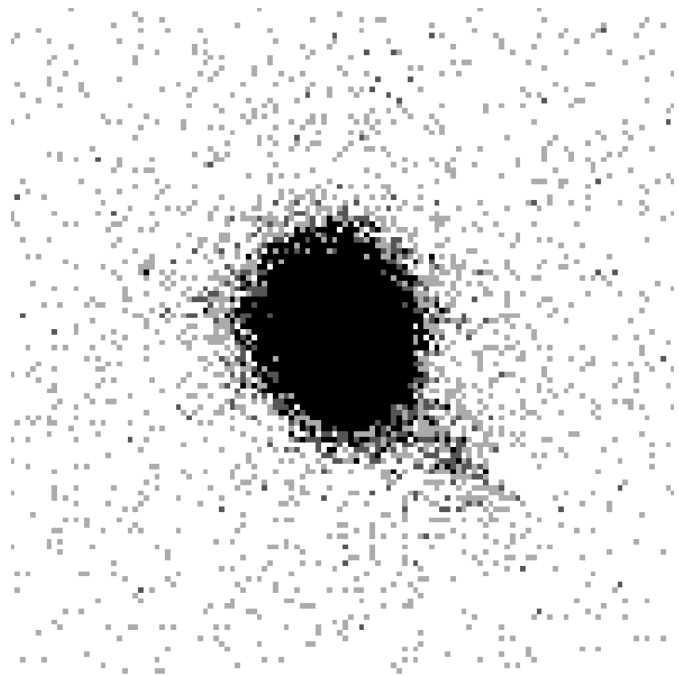}
\caption[]{\label{Fig2}Unprocessed image of 3C\,273 from the first observing
block clearly showing the jet towards the SW.}
\end{figure}

\section{ROSAT Observations}
The data were collected in two observing cycles. In a 17.991\,ksec integration
(January 1992) we clearly detected the jet without any image processing (see
Figure\,\ref{Fig2}). As the signal-to-noise ratio (S/N) was not sufficient for
detailed studies, 3C\,273 was re-observed in December 1994/January 1995 for a
total of 68.154\,ksec (quoted times as ``accepted'' by the ROSAT standard
reduction analysis).

\subsection{Improving the resolution}

\begin{figure}[htbp]
\centering
\includegraphics[width=8.8cm,clip=true]{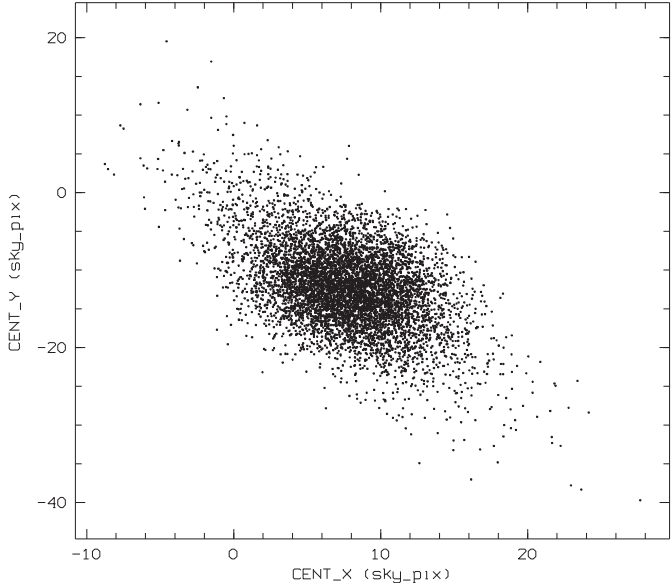}
\caption[]{\label{Fig3}Centroids of the quasar photons for all 10sec intervals
of observing block 2.}
\end{figure}

Images (sky pixel size 0\farcs5) have been produced from the event tables using
the EXSAS package provided from the ROSAT group at MPE in Garching. However,
the resulting point-spread function turned out to be unsatisfactory. Whereas in
the call for proposals the integral point response function of the ROSAT-XTE +
HRI was quoted to have a width of 5\arcsec, the FWHM of the images as produced
from the raw data was typically larger than 6\arcsec. This discrepancy is due
to inaccuracies in the aspect solution, which determines the sky coordinates
for each photon detected. Obviously in the standard aspect solution the
spacecraft wobble with a period of 402\,sec is not completely corrected for
(see Figure\,\ref{Fig3}).

\begin{table*}
\begin{tabular}{cccccc}
    \textbf{HZ 43} & \textbf{$\Delta $t} & \textbf{FWHM 1} &
    \textbf{FWHM 2} & \textbf{FWHM1 / } & \textbf{Max1 /}\\
    \textbf{data set \#} &\textbf{[sec]}&\textbf{[sky pixel]}&\textbf{[sky
    pixel]}&\multicolumn{1}{c}{\textbf{FWHM2}}&\multicolumn{1}{c}{\textbf{Max2}}\\
    100194&5806&7.64&15.44&0.49&1.88\\
    141873&9142&8.40&18.53&0.45&5.57\\
    142544&2719&9.11&18.93&0.48&4.63\\
    142545&2569&8.81&19.34&0.46&5.48\\
    142546&5353&9.48&20.03&0.47&5.47\\
    142547&2275&8.67&19.17&0.45&5.50\\
    142549&5884&9.17&19.71&0.47&5.36\\
    142550&2608&8.62&18.56&0.46&4.59\\
    &average& 8.89 & 19.18&&\\
    &r.m.s.&0.37&0.56&&\\
    \textbf{3C\,273 data}&85067& 9.065 &19.00 & 0.477&\\
\end{tabular}
\caption[]{\label{Tab1}Comparison of the point-spread-function for the white
dwarf HZ\,43 and 3C\,273 in terms of two Gaussian components with widths FWHM1
and FWHM2 (see Figure\,5). Units are sky pixels of 0 \farcs5.}
\end{table*}

With a count rate in the HRI of 2.8\,cts/sec the signal from the quasar core
itself is sufficiently strong to allow a shift-and-add procedure as follows:
All integrations are divided into time bins of 10\,sec duration and the centre
of gravity of the photons from the quasar core are calculated for each
interval. Photons were restricted to a raw amplitude between 2 and 8, typical
for the quasar. The average accuracy of the centroid positions was 0\farcs64
in X and 0\farcs70 in Y-direction. The interval of 10\,sec was a compromise
between sufficient time resolution and positional accuracy. These offsets from
the nominal centre position as a function of time directly correspond to the
remnant pointing errors due to the insufficiently corrected wobble motion.
They were interpolated in time by splines and the detected position of every
photon was corrected as a function of its arrival time.

\begin{figure}[htbp]
\centering
\includegraphics*[width=8.8cm,clip=true]{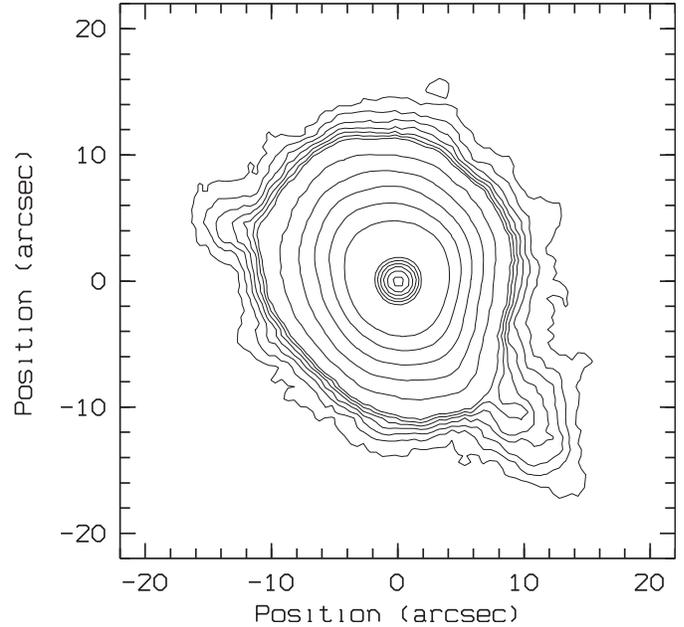}
\caption[]{\label{Fig4}Isophote plot of the final image (left). Contours are
at 2, 3, \ldots\ 8, 16 \ldots\ 1024, 1124, \ldots\ 1824 counts. Note the jet to
the SW and the source to the NE of the quasar. An elongation of the isophotes
in the general direction of the jet is evident.}
\end{figure}

\begin{figure}[htbp]
\centering
\includegraphics[width=8.8cm,clip=true]{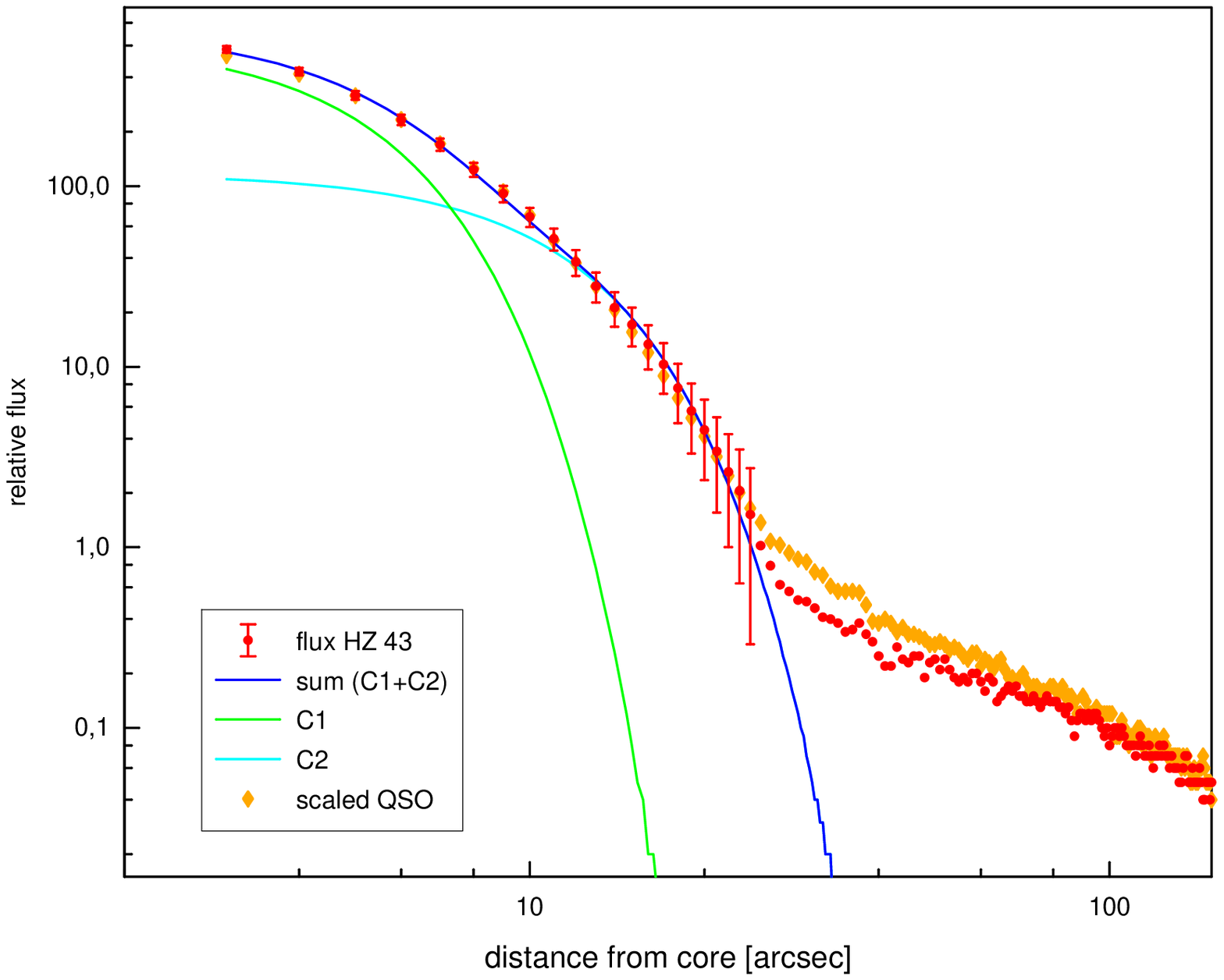}
\caption[]{\label{Fig5}Radial profile of 3C\,273 compared with that of the
white dwarf HZ\,43 (observation \#141873) after re-centring of the photons
(refer also to Table\,\ref{Tab1}). Not the good coincidence in the inner part
with the sum of two Gaussians for both objects. Error bars have been omitted
beyond 24\,pixels to bring out the difference bewtween white dwarf and quasar
halo.}
\end{figure}

\begin{figure}[htbp]
\centering
\includegraphics[width=8.8cm,clip=true]{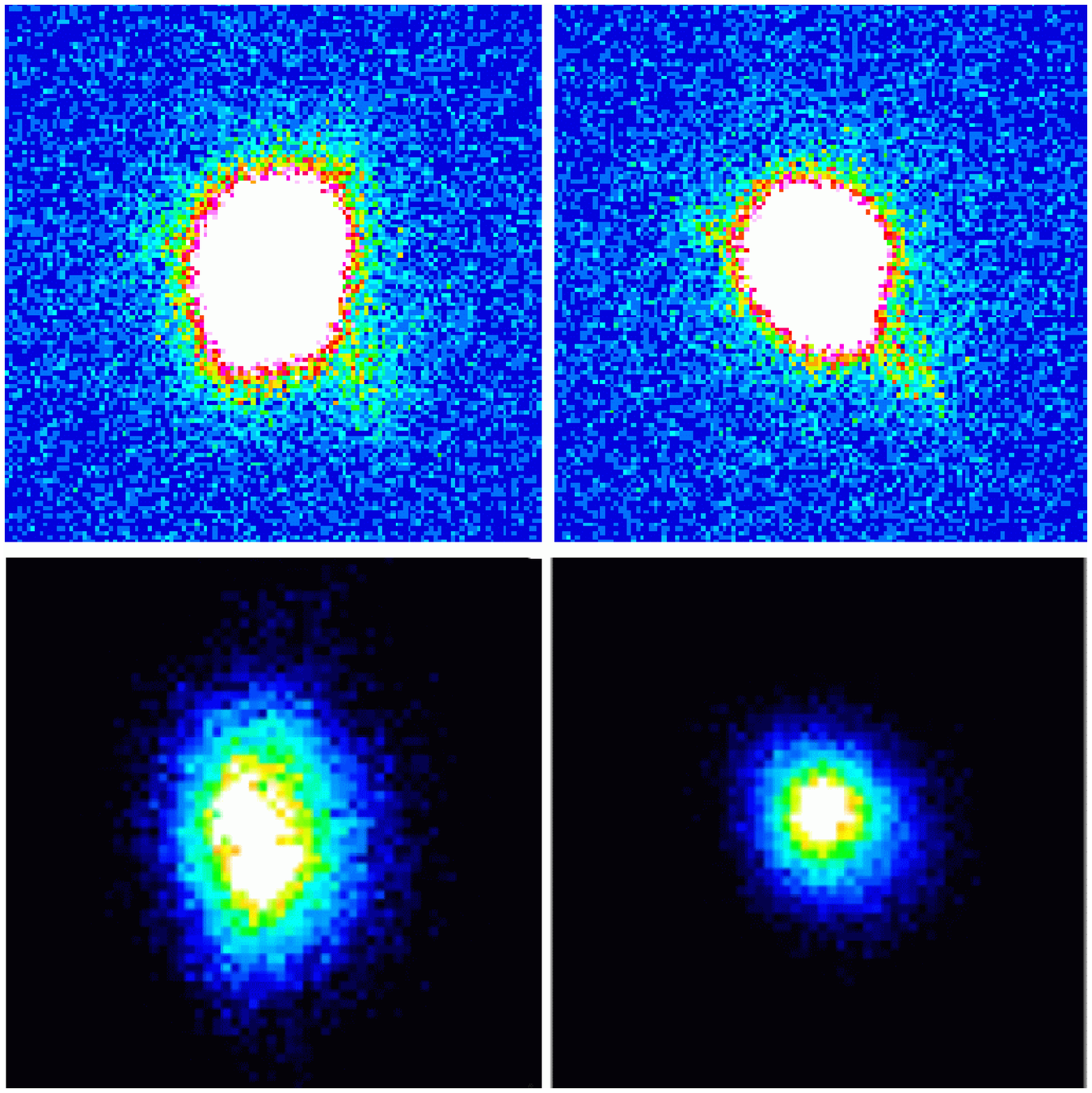}
\caption[]{\label{Fig6}ROSAT HRI image of 3C\,273 from the second observing
block before and after re-centring the photons (top). Corresponding images of
HZ 43 (\#142544) are shown for comparison (bottom). The QSO images
are optimised to show the jet, whereas the white dwarf images should best show
the circularisation of the PSF.}
\end{figure}

For comparison the same procedure was applied to eight data sets from the
ROSAT archive for the white dwarf HZ\,43, which is definitively a point source.
To derive the radial image profile, each data set was sampled on concentric
circles around the quasar respectively white dwarf with radii in steps of
0\farcs5. For each circle a constant was fitted to the data giving the
azimuthally averaged intensity profile. The ``core'' of these profiles was
decomposed by a least-square-fitting procedure into two Gaussians, neglecting
the very extended exponential component discussed in the HRI calibration
report (David \etal\cite{DHK99}). The shape of the point-spread-function (PSF)
in the raw images changed from observation to observation. But as can be seen
from Table\,\ref{Tab1} the profile of the de-jittered images was constant
within the error. As indicated in Table\,\ref{Tab1} and demonstrated by
Figure\,\ref{Fig5} and Figure\,\ref{Fig6}, the resolution could be enhanced to
4\farcs5 this way. A similar procedure was described by Morse (\cite{Morse94}).

From this analysis we conclude that there is no discernable difference in the
core profile between quasar and white dwarf in the averaged intensity
profiles, \emph{i.e.} the quasar core is unresolved at X-rays. Only in the
halo the intensity of the quasar is slightly above the normalized white dwarf
profile.

\subsection{Fitting the quasar's point-spread-function}

\begin{figure*}[htbp]
\centering{
\includegraphics[width=8.8cm,clip=true]{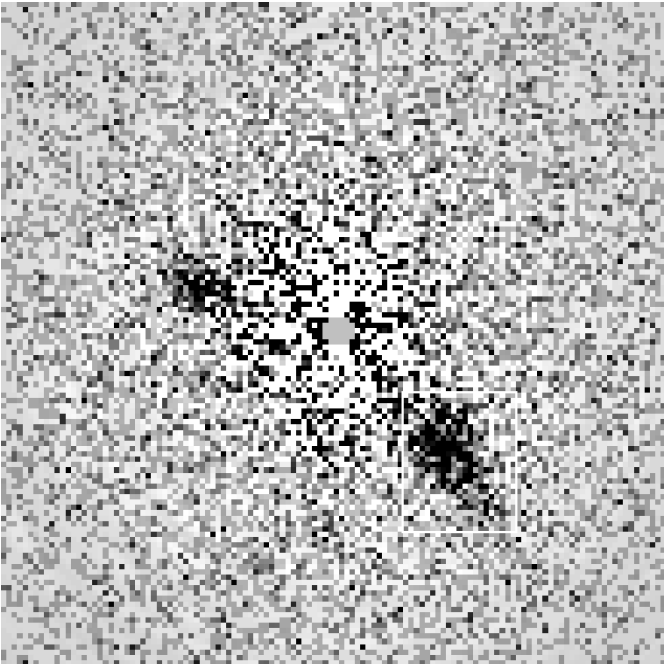}\hspace{\fill}
\includegraphics[width=8.8cm,clip=true]{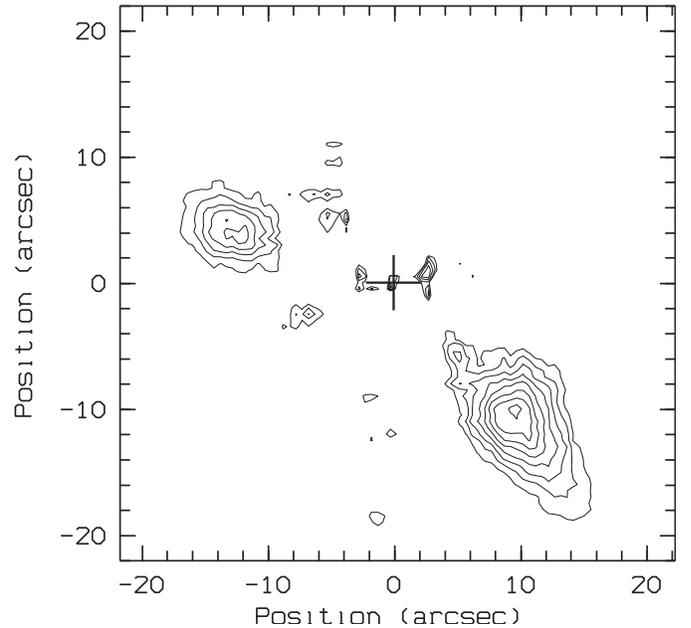}}
\caption[]{\label{Fig7}Background subtracted image of 3C\,273. The background
underneath the quasar is flat and shows the increased noise due to the intense
signal subtracted. The white rectangle indicates the region over which the
jet's X-ray emission was summed up. The isophotes clearly show the extent of
the X-ray emission all along the jet.}
\end{figure*}

\begin{figure}[htbp]
\centering
\includegraphics[width=8.8cm,clip=true]{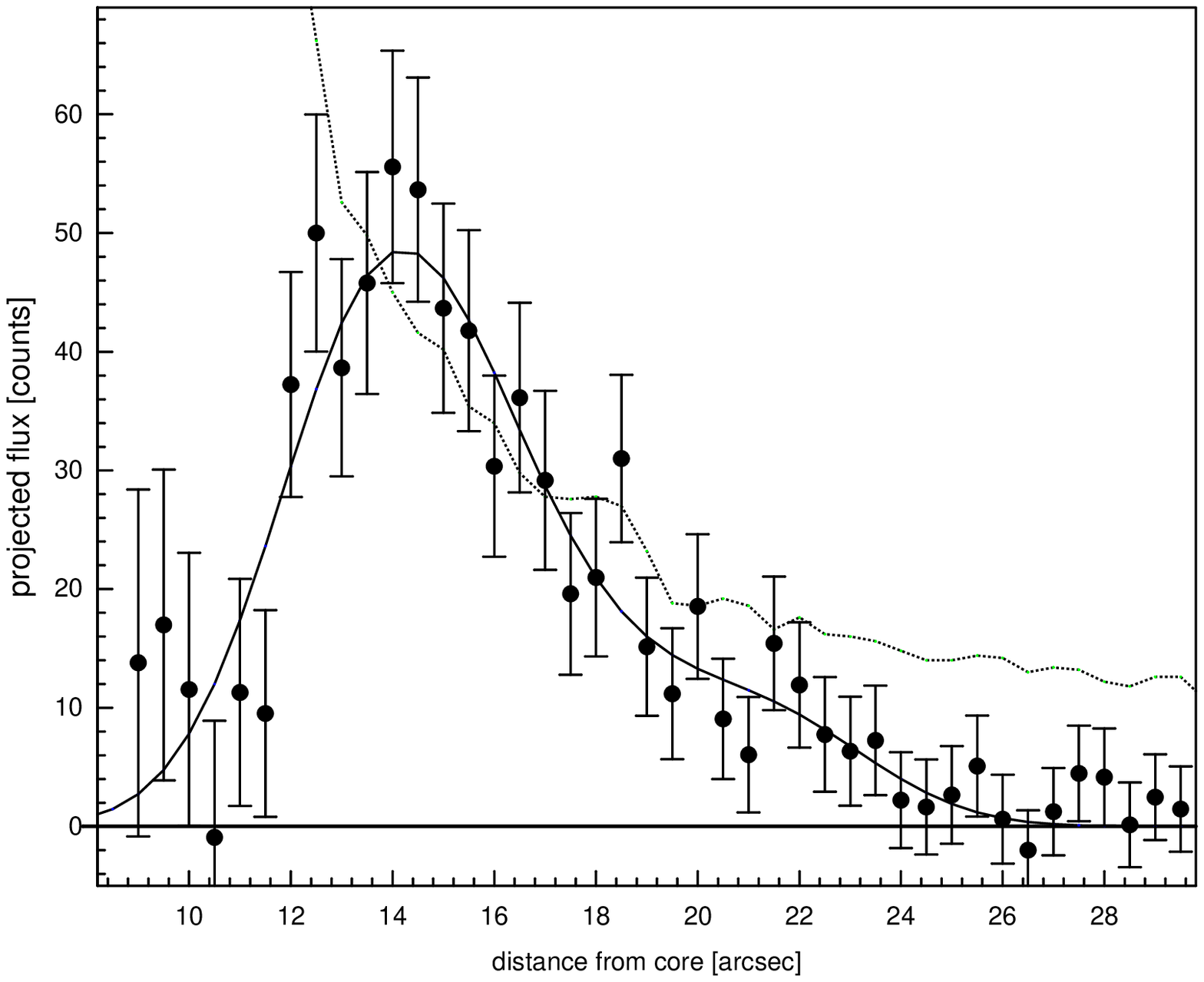}
\caption[]{\label{Fig8}Trace along the jet. The thin line represents the
background signal due to the quasar, summed over the same area as the jet
signal. The thick line is the model fit by 5 Gaussians described in the text.}
\end{figure}

The X-ray signal from the jet of 3C\,273 is very weak and is located in the
wings of the complicated point-spread-function of the bright quasar core (total
intensity ratio 400:1). Therefore the X-ray emission from the jet has to be
isolated from this underlying point-spread-function background. Using the
azimuthally averaged profiles derived above to analyse the
point-spread-function is not sufficiently accurate to isolate the jet's weak
signal and measure its flux. A better model of the point-spread-function had
to be obtained by fitting the signal sampled along concentric circles in four
sections by polynomials of order 3, with section boundaries at 30, 120, 210
and 300$^\circ$. In this way a satisfactory flat background also interior to
the optically visible jet region is achieved (Figure\,\ref{Fig7}). It should be
pointed out that the result of the background modelling does strongly depend
on the model parameters. With the current resolution of 4\farcs5 and the steep
intensity gradients towards the quasar core it cannot be completely excluded
that the relatively complicated model absorbs a moderately extended component
in the inner part of the jet ( $r < 10\arcsec$).

\subsection{The X-ray signal from the jet}
The count rate of the X-ray emission from the jet was derived from the
PSF-subtracted image. In a window encompassing the jet we measured 644\,counts.
The background in several windows of the same size and at the similar distance
to the quasar was measured to be $(28.6{\pm}13.5)$\,counts. We therefore deduce a
total jet flux of 610\,counts in 85068\,sec. The same procedure yields
253\,counts for the object to the NE of the quasar. To convert this to a flux
density we integrated a synchrotron power-law \emph{f}$_{\nu } \propto
$\emph{$\nu $}$^{\alpha }$ over the effective collecting area of the ROSAT HRI
as a function of frequency (effective energy 1.17\,keV corresponding to
$2.83\times10^{17}$\,Hz) including absorption due to a neutral hydrogen column
of $1.8\times10^{20}$\,cm$^{-2}$ (Otterbein \cite{Otterbein92}) using the
cross-sections from Morrison \& McCammon (\cite{MMcC83}) for a solar
abundance. To represent synchrotron emission we used spectral indices $\alpha
= -1$ and \quad $-2$ and thus obtained an integral flux density of $(36.6{\pm}2.8)$
and $(14.9{\pm}1.1)$\,nJy for the jet and $(15.2{\pm}1.1)$ and $(6.2{\pm}0.5)$\,nJy for the
object in the NE of the quasar. If we assume a spectral index of $\alpha = 0$
(resembling thermal bremsstrahlung) the corresponding values are
$(57.1{\pm}4.3)$\,nJy for the jet and $(23.7{\pm}1.8)$\,nJy for the object in the NE.
The error was formally derived from the scatter in the background
determination.

\begin{table}
\begin{tabular}{|c|c|c|c|c|}
\hline \raisebox{-1.50ex}[0cm][0cm]{Knot}&
\raisebox{-1.50ex}[0cm][0cm]{Position \par [\arcsec]}&
\multicolumn{3}{|c|}{Flux [nJy]}\\
\cline{3-5}
 & & $\alpha = - 2$& $\alpha = -1$&$\alpha = 0$ \\
\hline A & 13.2 & 5.9 & 14.4 & 22.5 \\
\hline B & 15.0 & 4.7 & 11.4 & 17.8 \\
\hline C & 17.2 & 1.9 &  4.8 &  7.5 \\
\hline D & 19.8 & 1.2 &  2.9 &  4.6 \\
\hline H & 22.0 & 1.2 &  3.0 &  4.7 \\
\hline \multicolumn{2}{|c|}{sum} & $14.9{\pm}1.1$& $36.6{\pm}2.8$&
$57.1{\pm}4.3$ \\
\hline
\end{tabular}
\caption[]{\label{Tab2}X-ray flux of the jet's components in the ROSAT HRI
band (at 1.17\,keV corresponding to $2.83\times10^{17}$\,Hz).}
\end{table}

\begin{figure}[htbp]
\centering
\includegraphics[width=8.8cm,clip=true]{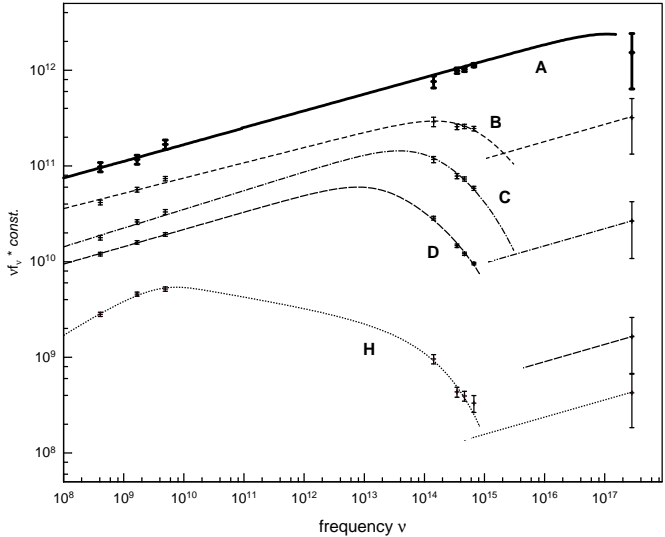}
\caption[]{\label{Fig9}Spectra of the individual knots along the jet of
3C\,273. The spectra of the individual knots were multiplied by constants to
shift them vertically for clarity (0.5 for H, lowest spectrum, 2 for D, 20 for
C, 100 for B and 380 for A, uppermost spectrum). The X-ray points at the far
right cover the range found for the range in spectral index between 0 and $-2$.
These points cannot be a continuation of the synchrotron spectra except
possibly for knot\,A. The power-law meeting the X-ray point (except for
knot\,A) indicates a hypothetical second synchrotron component with a spectral
index of $-0.82$ (low frequency average over knots A to D). Such a component
would not be detectable in currently observed wavebands except the X-rays.}
\end{figure}

Inspection of Figure\,\ref{Fig7} suggests that the X-ray emission from the jet
is extended in the radial direction all along the visible jet. We therefore
rotated the background subtracted image by 47\fdg8 around the quasar core to
sum the jet's signal over the 20 pixel rows encompassing it. The resulting
trace along the jet is shown in Figure\,\ref{Fig8}. This run of the X-ray
emission along the jet was analysed in a similar manner as the optical and
radio emission (R\"{o}ser \& Meisenheimer \cite{HJRM91}): Gaussian profiles
with constant width of 4\farcs5 were placed at the positions of knots A, B, C,
D and H\footnote{Nomenclature introduced by Leli\'{e}vre
\etal(\cite{LNHRS84}).}. Their peak was adjusted via a least square fit to
represent the trace of X-ray emission along the jet (Table \,\ref{Tab2}).

The strongest X-ray signal originates from the positions of knots\,A and B,
much weaker emission is found further out, possibly out to the hotspot H. Due
to the steeply rising quasar background, and to uncertainties in background
modelling mentioned above, the onset of the jet's X-ray emission is somewhat
uncertain. On the basis of our data no X-ray emission is detected from the jet
inwards of knot\,A.

The X-ray source to the NE of the quasar does not show any optical counterpart
on our deep images, neither in the radio nor in the optical. Due to its
weakness we cannot say if it is extended or not in the X-rays. Its relation to
3C\,273 and its nature remain unknown.

\section{Origin of the X-ray emission}
To shed light onto the possible emission mechanism(s) giving rise to the jet's
X-ray emission, the above data were combined with results from our
radio-to-optical studies at 1\farcs3 resolution (Meisenheimer
\etal\cite{MYHJR97}, Neumann \etal\cite{NMHJR97}, R\"{o}ser \& Meisenheimer
\cite{HJRM91}) to investigate the run of the continuum across the
electromagnetic spectrum for the different regions along the jet. We plot the
spectra of knots A, B, C, D, and H in Figure\,\ref{Fig9}, where the flux was
multiplied by different factors to disentangle the individual spectra in the
graph. The continua are model spectra as described by Heavens \& Meisenheimer
(\cite{HM87}) and Meisenheimer \& Heavens (\cite{MH86}). They result from Fermi
acceleration of particles in a strong shock and include the radiation losses
in a finite down-stream region. The exponential cut-off reflects the maximum
energy obtained by the particles and is well established for knots B to H also
by our recent HST WFPC2 data at 300\,nm (Jester \emph{et al.}, in
preparation). It is evident that in general the X-ray flux level is not a
continuation of the radio-optical synchrotron cut-off continuum. Therefore
different emission mechanisms have to be discussed for the individual knots.

\subsection{Synchrotron radiation from the jet}
Only for knot\,A does the extrapolation of the radio-to-optical continuum
approximately meet the X-ray flux level. Extrapolation of the $\lambda$6cm
flux with the low frequency spectral index of knot\,A ($\alpha   = - 0.83)$
into the ROSAT range predicts an X-ray flux of 32\,nJy, only a factor of
roughly two above the observed level. Whereas in Meisenheimer
\etal(\cite{MNHJR96p}) a lower limit to the cut-off frequency for knot\,A
could be set at about $5\times10^{16}$\,Hz, we now have to increase this value
by about a factor of 50 (see Figure\,\ref{Fig9}). According to standard
synchrotron theory we can therefore place a new lower limit to the maximum
energy of the relativistic particles in knot\,A of
\[\gamma _{c} = 10^{6} \times \sqrt {{{\frac{{\nu _{c} }}{{1.26 \times
10^{15}{\kern 1pt} Hz}}} \mathord{\left/ {\vphantom {{\frac{{\nu _{c} }}{{1.26
\times 10^{15}{\kern 1pt} Hz}}} {\frac{{B_{ \bot } }}{{30{\kern 1pt} nT}}}}}
\right. \kern-\nulldelimiterspace} {\frac{{B_{ \bot } }}{{30{\kern 1pt}
nT}}}}} = 3 \times 10^{7}\]

\noindent where we have used the minimum energy field of 67\,nT. For the other
knots the primary synchrotron component producing the observed radio-to-optical
continuum exhibits an exponential cut-off in the optical/UV-range
(Meisenheimer \etal\cite{MNHJR96p}). To maintain the synchrotron scenario also
for these regions a second particle population with higher maximum energy has
to be invoked\footnote{Such a second synchrotron component was also suggested
by Harris \etal\ (\cite{HHSSV99}) to account for the X-ray emission from the
jet in 3C\,120.}. These populations are indicated in Figure\,\ref{Fig9}
connecting to the ROSAT HRI points with an assumed spectral index of $\alpha
=   - 0.82$, the low-frequency average for knots A to D. The required density
of relativistic particles in these hypothetical components decreases outwards
along the jet. It is a factor of 5 below the density of the particles
producing the observed radio-to-optical continuum for knot\,B and a factor of
15 and 200 below that in knot\,C and D, respectively. The small fraction of
relativistic particles make it very hard to detect them at \emph{e.g. }optical
wavelengths. However, if this second population is confined to some centres of
very effective acceleration on sub-arcsecond scales, we expect to see
deviations from the cut-off spectrum in the optical spectral index map derived
from our HST R- and U-band data at a resolution of 0\farcs2 (Jester \emph{et
al.} 2000, in preparation).

\subsection{Synchrotron Self-Compton emission (SSC) from the jet}
Prime sources for inverse Compton emission are compact regions with high radio
photon densities in the jet, for which the hot spot H is the most likely
candidate. The size of the emission region and the radio flux originating from
it determine the amount of synchrotron-self-Compton emission. We have
calculated the inverse Compton emission along the lines given by Blumenthal \&
Tucker (\cite{BT70}) as follows: The most compact component certainly is the
hot spot's acceleration region, where the optically radiating particles are
produced. Its contribution to the synchrotron spectrum can be inferred from
the low-frequency power-law and the high-frequency cut-off. The intermediate
range with the break in the continuum (see Figure\,\ref{Fig9}) is dominated by
the superposition of the down-stream regions, where the relativistic particles
already have lost part of their energy. Connecting the cut-off part smoothly
with a power-law of index $-0.60$ we estimate the 5\,GHz-flux from the
acceleration region itself to be about 0.1\,Jy. From the most recent analysis
of the hot spot's spectrum by Meisenheimer \etal(\cite{MYHJR97}) we infer a
thickness of the emission region close to the internal shock (Mach disk) of
1.4\,pc, the width perpendicular to the jet (from our best-resolved radio map)
is taken to be 2.2\,kpc (circular cross-section assumed). For a minimum energy
field of 35\,nT (assumed constant over the entire hot spot in the model) we
set the density in relativistic particles in this volume to reproduce the
above estimated $\lambda $6cm flux of 0.1\,Jy. Integrating this synchrotron
emission over the range 10\,MHz to $5\times10^{4}$\,GHz (corresponding to
Lorentz-factors of 100 to $2\times10^{5}$) produces an inverse Compton flux of
about 1\,nJy, to within factors of 2 to 4 what is observed (see
Table\,\ref{Tab2}). As this is only a rough estimate to check the order of
magnitude the discrepancy could well be removed by fine-tuning the parameter
assumed (filling factor, geometry…).

For the other knots synchrotron-self-Compton emission fails to meet the
observed level by orders of magnitude, \emph{e.g.} for knot\,A we expect an
inverse Compton X-ray flux of only 0.01\,nJy. As the photon density of the
microwave background radiation is one order of magnitude less than the
synchrotron photon density in all knots, its photons cannot account for the
X-ray flux via inverse Compton scattering either.

\subsection{Thermal Bremsstrahlung from the jet}
If future high resolution observations fail to reveal locations of $\nu _{c}
\gg 10^{15}$\,Hz in knots B, C, and D, there remains the bremsstrahlung
emission from a thermal plasma as a last resort to explain the jet's X-ray
emission outside knot\,A and the hot spot. A plasma at 10$^{8}$\,K and with an
electron density of 1\,cm$^{ - 3}$ spread over the volume of a typical jet
knot ($0\farcs5\times0\farcs2$ from our HST images, Jester \emph{et al.} 2000,
\emph{in preparation}) would produce an X-ray flux in the ROSAT window
corresponding to 0.01\,nJy. Even with this unrealistically high electron
density this is orders of magnitude below the observed X-ray flux level.
Furthermore any sufficiently dense thermal plasma would result in total
depolarisation of the jet's radio emission and in a substantial rotation
measure. Both are not observed (Conway \etal\cite{CGPB93}). So thermal
bremsstrahlung is highly unlikely to account for the observed X-ray flux.

\section{The X-ray halo of 3C\,273}
\begin{figure}[htbp]
\centering
\includegraphics[width=8.8cm,clip=true]{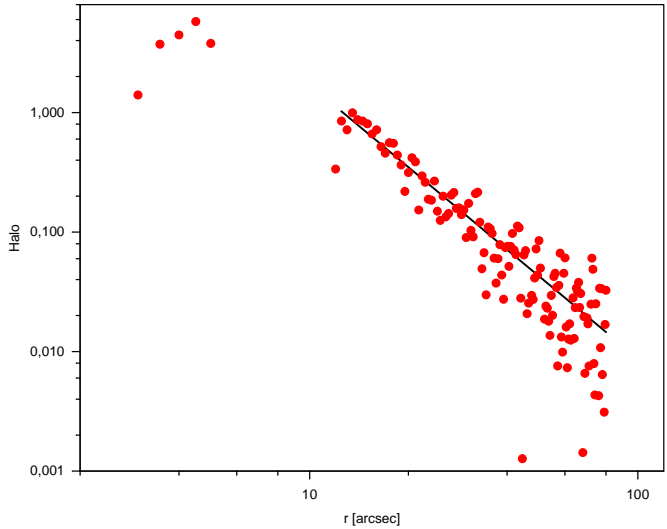}
\caption[]{\label{Fig10}Azimuthally averaged radial profile of the quasar
halo's X-ray emission. The fit with a King profile is shown between 12\farcs5
and 80\arcsec\ from the core.}
\end{figure}

For the comparison of the quasar's profile with that of the white dwarf, the
normalization constant had been optimised over the range 3\arcsec\ to
13\arcsec, where the profiles show excellent agreement. However, it is evident
from Figure\,\ref{Fig5} that the quasar profile lies systematically above the
profile of the stellar source beyond radii of about 15\arcsec. Although the
error of individual data points is large, we regard the systematic deviation
as real and attribute it to an extended X-ray halo of the quasar (3C\,273 had
been observed on-axis, the pointing for HZ\,43 was 1\arcmin\ off-axis). To
isolate the X-ray flux from this halo we have used the scaling from
Figure\,\ref{Fig5} and subtracted the azimuthally averaged HZ\,43-profile from
the averaged quasar profile (Figure\,\ref{Fig10}). This differential profile
was fitted with a King profile (Jones \& Forman \cite{JF84})

\[P(r) = P(0)\left[ {1 + \left( {r/a} \right)^{2}} \right]^{ - 3*\beta + 0.5}\]

\noindent with \emph{P}(0) = 3.6, \emph{a} = 10\farcs8 $\hat{=}$ 28.8\,kpc,
\emph{$\beta $} = 0.6758 determined from a least square fit to the data points
between 12\farcs5 and 80\arcsec\ from the core. Although the formal errors for
these parameters are large, we nevertheless used them for our numerical
estimates, as they describe the data reasonably well. The King-profile was
integrated up to give the total X-ray flux from the halo of 232\,nJy
corresponding to a luminosity in the HRI band of $3.4\times10^{43}$\,erg/sec
(spectral index $\alpha   =  0$ assumed). To convert this into an estimate of
the physical parameters we assumed a plasma temperature of
$2.7\times10^{7}$\,K as in the M\,87 halo (B\"{o}hringer \etal\cite{BBSVHT94}) and
that the plasma is confined with constant density to within the core radius.
From these assumptions a density of $6\times10^{4}$ H-atoms/m$^{3}$ is derived
for the central part of the halo. This is a factor of 10 above the upper limit
derived from EINSTEIN observations by Willingale as quoted by Conway
\etal(\cite{CDFR81}). The typical density of the thermal plasma in the cores
of rich clusters of galaxies is a factor of 10 below this value (Jones \&
Forman \cite{JF84}) but the density derived above for the halo of 3C\,273
matches the central density of $(7{\pm}2)\times10^{4}$\,m$^{ - 3 }$ derived for the
central density derived for the intra-cluster medium around Cygnus\,A (Carilli
\etal\cite{CPH94}).

\section{Conclusions}
ROSAT HRI observations confirm the X-ray emission from the jet of 3C\,273 as
suspected on the basis of the EINSTEIN HRI observations. They furthermore show
that the X-rays originate from all along the jet, probably including the hot
spot. While the jet's radio emission is highly peaked towards the outer end,
the optical emission is more or less constant along the jet. This trend with
wavelength continues at X-rays in that these are strongly peaked now at the
inner end (knots\,A/B).

Despite the considerably improved data base the problems with the
interpretation of the jet's X-ray emission (see Harris \& Stern \cite{HS87x})
still remain. Whereas the X-ray emission from the jet of M\,87 might well be
due to the synchrotron emission process (Neumann \etal\cite{NMHJRF97},
R\"{o}ser \& Meisenheimer \cite{RingbergIV}), the situation is less clear for
the jet of 3C\,273. An extrapolation of the radio-to-optical synchrotron
continuum could only explain the X-ray emission from the innermost knot\,A, it
fails for the rest of the jet. Any X-ray emission from the hot spot up to the
level we found can be accounted for by synchrotron-self-Compton emission. The
emission mechanism for knots\,B to D remains a mystery, as --- except for an
hypothetical high-energy synchrotron component --- all three mechanisms
discussed above cannot account for the X-ray emission.

The forthcoming observations of 3C\,273 by CHANDRA will provide X-ray data at
higher resolution and with spectral information. High spatial resolution will
test if the X-ray emission does indeed coincide with the radio-optical
synchrotron continuum emission, not necessarily the case for thermal
bremsstrahlung emission. X-ray spectra will set important constraints on the
emission mechanism mainly via the spectral slope in the X-ray band, which could
directly reveal a second population of relativistic particles emitting
synchrotron X-rays. Thus we can expect further insight into the mystery of
X-ray emission from the jet of 3C\,273 within the immediate future.

\begin{acknowledgements}
We thank G. Hasinger for discussions on the ROSAT data reduction and his
suggestion to try re-centring the photons. Valuable advice from C. Izzo and S.
D\"{o}bereiner is also kindly acknowledged.
\end{acknowledgements}

\end{document}